\documentclass[useAMS,usenatbib]{mn2e}
\usepackage[]{txfonts}
\usepackage{color}
\def\simless{\mathbin{\lower 3pt\hbox
{$\rlap{\raise 5pt\hbox{$\char'074$}}\mathchar"7218$}}}   
\def\simmore{\mathbin{\lower 3pt\hbox
{$\rlap{\raise 5pt\hbox{$\char'076$}}\mathchar"7218$}}}   
\newcommand{\be}{\begin{equation}}
\newcommand{\ee}{\end{equation}}
\usepackage{graphicx}
\usepackage{epsfig} 
\usepackage{epstopdf}

\newcommand{\fig}[1]{Fig.~\ref{fig:#1}}
\newcommand{\fign}[1]{\ref{fig:#1}}
\newcommand{\eq}[1]{Eq.~(\ref{eq:#1})}
\def\comp{\,c/\omega_{\rm p}}
\newcommand{\eqb}{\begin{eqnarray}}
\newcommand{\eqe}{\end{eqnarray}}

\title[Relativistic jets shine through shocks or magnetic reconnection?]{Relativistic jets shine through shocks or magnetic reconnection?}

\author[Lorenzo Sironi, Maria Petropoulou and Dimitrios Giannios]
{Lorenzo Sironi,$^1$\thanks{NASA Einstein Post-Doctoral Fellow. E-mail: lsironi@cfa.harvard.edu.} 
Maria Petropoulou$^{2}$\thanks{NASA Einstein Post-Doctoral Fellow.} and Dimitrios Giannios$^{2}$\\
$^1$Harvard-Smithsonian Center for Astrophysics, 
60 Garden Street, MA 02138, USA\\
$^{2}$Department of Physics and Astronomy, Purdue University, 525 Northwestern
Avenue, West Lafayette, IN 47907, USA}

\begin{document}
\date{Received / Accepted}
\pagerange{\pageref{firstpage}--\pageref{lastpage}} \pubyear{2013}

\maketitle

\label{firstpage}

\begin{abstract}
Observations of gamma-ray-bursts and jets from active galactic nuclei reveal that the jet flow is characterized by
a high radiative efficiency and that the dissipative mechanism
must be a powerful accelerator of non-thermal particles. Shocks and
 magnetic reconnection have long been considered as
possible candidates for powering the jet emission. Recent progress via fully-kinetic particle-in-cell simulations allows us to revisit this
issue on firm physical grounds. We show that shock models are unlikely to account for
the jet emission. In fact, when shocks are efficient at dissipating energy,
they typically do not accelerate particles far beyond the thermal energy, and vice
versa. In contrast, we show that magnetic reconnection can
deposit more than 50\% of the dissipated energy into non-thermal leptons
as long as the energy density of the magnetic field in the bulk
  flow is larger than the rest mass energy density. The emitting region, i.e., the reconnection downstream,
is characterized by a rough energy equipartition between magnetic fields and radiating particles,
which naturally accounts for a commonly observed property of blazar jets.
\end{abstract} 
  

\section{Introduction} 
\label{intro}
{Relativistic jets are ubiquitous in the Universe, with gamma-ray bursts (GRBs)
and active galactic nuclei (AGNs) being two representative examples of high-energy
sources powered by jets.}
Despite decades of research, the issue of
{what powers}
relativistic astrophysical jets is still unresolved.
On the one hand, there is a strong theoretical motivation to
consider them as magnetically-dominated objects at their base
\citep{blandford_77,blandford_82}. The strong magnetic fields threading a rotating compact object or the associated accretion disk serve to convert the rotational energy of the central engine into the power of the outflow.
Part of the magnetic energy is used to accelerate the flow to relativistic speeds but, 
generally, magnetohydrodynamic (MHD) models predict that the largest fraction of the energy remains locked in the magnetic field,
i.e., the jet arrives Poynting-dominated at the dissipation distance
\citep{sasha_09,komissarov_09,lyubarsky_09}. On the other hand, {leptonic}
models of the radiative signature of {blazars} -- a subclass of AGNs whose jets point along our line of sight --  indicate that the 
energy densities of magnetic field $U_{\rm B}$ and radiating particles  $U_{\rm e}$ at the
emission region do not differ by more than one order of magnitude \citep{readhead_94,celotti_08,ghisellini_14}.

Our ignorance on the {source of jet power} 
allows for two very different scenarios {regarding} the dissipative mechanism behind the jet emission. Hydrodynamical flows can dissipate
energy and efficiently accelerate particles at shock fronts 
(e.g., \citealt{heavensdrury_88, rees_94}), {whereas in strongly magnetized flows the efficiency of shocks is greatly
reduced \citep{kennel_84}. In this case, a more likely candidate  for powering the jet emission is the process of 
magnetic reconnection  \citep{spruit_01,spruit_02}, 
where the annihilation of field lines of opposite polarity transfers the field energy to the particles.}

{Observations of relativistic jets 
can be used to infer the properties of the emitting region.
Yet, these} do not directly represent the typical conditions in 
 the bulk flow. Emitting regions are, by definition, special places
were intense energy dissipation has occurred. {Both} the shock
downstream and the reconnection outflow qualify as such regions. {On the contrary,
the upstream medium (in the case of shock dissipation) or the inflow region (for the case of reconnection)
describe more closely} the bulk of the jet. 
{In this sense, the observational inference 
on the ratio $U_{\rm B}/U_{\rm e}$, 
applies directly to the emitting region and not to the bulk of the jet flow. Hereafter, we identify the (shock or
reconnection) upstream with the large-scale jet and the downstream 
with the emitting region.} 

{The properties of the emitting region are related to those of the bulk of the jet in a way
that depends on the dissipative process. In shocks}
the magnetic field is generally compressed and {its strength is, thus,}
increased: the magnetic energy per particle is larger in the downstream 
than in the upstream.\footnote{This condition is verified in quasi-perpendicular  
shocks (i.e., where the field lines are nearly orthogonal to the shock direction of propagation), 
which is the most relevant field configuration for relativistic jets, due to Lorentz transformation effects and shock compression.} 
In contrast {to shocks}, magnetic reconnection dissipates magnetic energy: the downstream (i.e., the outflow) is 
less magnetized than the upstream (i.e., the inflow). 
In fact, the more efficient is reconnection in accelerating particles, 
the more particle-dominated the downstream region becomes. 

In both shocks and reconnection, the partition of energy between
magnetic fields and particles can be  described self-consistently 
via  fully-kinetic  particle-in-cell (PIC) simulations. Recent PIC progress has been remarkable. 
First-order Fermi acceleration in relativistic shocks has been demonstrated from first
principles
\citep{spitkovsky_08,spitkovsky_08b,martins_09,haugbolle_10,sironi_13}. 
However, as expected from theoretical arguments \citep{begelmankirk90}, no
particle acceleration takes place when relativistic shocks are
quasi-perpendicular, i.e., where the field lines are nearly orthogonal to the shock direction of propagation
\citep{gallant_92,sironi_spitkovsky_09,sironi_spitkovsky_11a}. Recent PIC simulations have also demonstrated that magnetic reconnection in magnetically-dominated plasmas
is a fast, unsteady process that leads to efficient particle acceleration \citep[e.g.,][for a review]{ss_14, kagan_15}. In both
shocks and reconnection, leptons pick up a large fraction of the
dissipated energy, so electrons are the most likely candidates for powering the observed emission (i.e., supporting the so-called leptonic scenario).

{In this work, we use jet observations 
to probe}
the nature of the dissipative mechanism and the properties 
of the bulk of the jet. For this, we exploit the latest PIC
 findings about shocks and magnetic reconnection as well as 
{three} basic facts about jet emission: (i) jets are efficient emitters,  
(ii) the radiating particles have extended, non-thermal  distributions, {and (iii) the 
energy densities stored in the magnetic field and in radiating particles do not differ more than one order
of magnitude.\footnote{We remark that the observational inferences (i) and (iii) are model-dependent, relying on the assumption of a leptonic origin for the observed emission. In this study, 
our working hypothesis is that the radiating particles are leptons, and our 
results should be interpreted in this framework.}
The paper is structured as follows.
In Section 2 we summarize these basic observational
constraints. In Section 3 we apply them to two alternative dissipative
scenarios: shocks and magnetic reconnection. We conclude in Section 4.}

\section{Inferences from Jet observations}
\label{obs}
{There is a consensus that relativistic jets} are efficient emitters. 
GRBs have efficiency $f\sim 0.1-1$
in converting jet power into {gamma-ray} (prompt)
emission \citep[e.g.,][]{berger00,panaitescu02,yost03,granot_06,nava_13}. 
{Blazars} have similarly high radiative efficiencies.
{Recently,} \citet{ghisellini_14} modelled the spectral energy distribution (SED) of a large sample of blazars
and, assuming one proton per radiating electron, they inferred a radiative efficiency $f\sim 0.1$ \citep[see also][]{celotti_08}. 
{The efficiency may be even higher, if blazar jets are pair-loaded, namely they contain more leptons than protons} \citep{ghisellini_14}. 
It follows that the dissipative mechanism behind
the jet emission must be able to transfer a large amount of energy to the radiating particles, achieving
a high radiative efficiency ($f\gtrsim 0.1$). {Given that $f$ is already
high, the dissipation process must involve a large fraction of the jet fluid, in order to explain the observed photon luminosities without
pushing the jet energetics to extreme values.}

{Multi-wavelength jet observations typically suggest that
the radiating particles are accelerated into a non-thermal power-law distribution.}
The blazar SED {has a characteristic double-humped shape (e.g., \citealt{ulrichetal_97, fossatietal_98}), with
a broad low-energy component extending from the radio up to the UV band, and in some extreme cases up to $\gtrsim 1$~keV X-rays \citep{costamante_01}. The high-energy component extends across the X-ray and $\gamma$-ray bands, with peak energy around
$0.1$ TeV, although this is not always clear (see, e.g., \citealt{abdo_11} for Mrk~421).
The low-energy hump is believed to result from synchrotron emission from relativistic electrons. Its spectral shape requires, in general, 
a power-law (or broken power-law) particle spectrum \citep[e.g.,][]{celotti_08}, which implies non-thermal 
particle acceleration.} Furthermore, the high-energy
hump of blazars, {which in leptonic scenarios is explained as inverse Compton emission,} 
extends in many sources up to 100 GeV or higher \citep[e.g., see][for PG~1553+113]{aleksic_12}. This also points
to electron acceleration up to Lorentz factors $\gamma_{\rm e} \gtrsim 10^5$, 
or even higher for TeV-emitting blazars.\footnote{In hadronic models for the gamma-ray emission from blazars, the required maximum Lorentz factor of protons
may be even more extreme \citep{aharonian_00, boettcher_13, petro_dim_pado_15}.} 
The same holds true in synchrotron models for the prompt emission of
GRBs \citep[and references therein]{daigne_11}.\footnote{Note however that the GRB prompt signature can be the result of photospheric emission from 
thermal electrons \citep{ghisellini_99,giannios_06,peer_06}.}  
In summary, the dissipative mechanism that operates in jets has to be able to 
accelerate particles well beyond their thermal energies.

In blazars both the synchrotron and the Compton humps are visible 
and their relative brightness constrains the ratio of energy densities 
$U_{\rm B}/U_{\rm e}$ at the emitting region {\citep[e.g.,][]{sikora_stawarz_09}. }The fact that in BL Lacs
the two bumps have comparable luminosity indicates that, in the 
synchrotron self-Compton (SSC) model, $U_{\rm B}/U_{\rm e}\sim 1$ {(see also \citealt{mastkirk_97}).
However, for gamma-ray luminous blazars, such as flat spectrum radio quasars (FSRQs), where the high-energy component
is much more luminous than the low-energy one, the SSC scenario would predict that $U_{\rm B}/U_{\rm e}\ll 1$, i.e.,
far from equipartition. Yet, a rough equipartition between the magnetic field and
radiating electrons is still inferred in such blazars, in the external Compton (EC) interpretation
of their gamma-ray emission \citep{sikora_stawarz_09}.}
In summary, detailed modeling of many blazars {within the SSC or EC scenarios} points to the fact that, quite generally, $U_{\rm B}/U_{\rm e}\sim 1$, but some emitting regions
may be moderately magnetically-dominated, i.e., $U_{\rm B}/U_{\rm e}\sim 3$
\citep{ghisellini_14}.  The rough equipartition between emitting
particles and magnetic field in the blazar emitting region is also suggested by the
observed surface brightness temperatures at radio wavelengths 
\citep{readhead_94,wilk_99,homan_06,hovatta_13}. In this work, we adopt the range  
$0.3\lesssim U_{\rm B}/U_{\rm e}\lesssim 3$ as representative of the 
physical conditions in the emission region of blazars \citep{coppi_99,celotti_08,ghisellini_14}.
In the case of GRBs, no {high-energy bump} in the SED can be securely associated with
the prompt emission, making the $U_{\rm B}/U_{\rm e}$ ratio at the emitting region 
harder to quantify.

In summary, any model for the jet emission 
has to account for the fact that
both blazars and GRBs are characterized 
by (i) a large radiative efficiency $f \gtrsim 0.1$ and by (ii)
extended non-thermal distributions of the radiating particles. In addition,  
 blazars are also characterized by (iii) a rough equipartition
of energy density between radiating electrons and magnetic field.


\section{Constraints on the dissipative mechanism}
\label{constraints}
{In this section we investigate whether the observational inferences -- on
the dissipative efficiency, the equipartition between particle and magnetic energies, and the requirement of non-thermal particle acceleration -- 
can constrain the dissipation mechanism responsible for the jet emission. Shocks and magnetic reconnection
are the most commonly invoked dissipation processes.} If the jet 
is dominated by kinetic energy flux at the
dissipation distance (low-$\sigma$ flow),\footnote{Throughout this work, the magnetization of the 
fluid is defined as $\sigma= B^2 / 4 \pi \rho c^2$, where both the magnetic field strength $B$ and
the mass density $\rho$ are measured in the rest frame of the fluid.} then shocks are natural candidates for  the required dissipation. 
{On the other hand, 
if the jet remains Poynting flux dominated at the dissipation distance,  magnetic reconnection 
is a more plausible candidate for powering the jet emission, since shock
dissipation is suppressed in high-$\sigma$ flows \citep{kennel_84,giannios_08,narayan_11}.
In this case, however, a process that triggers the reconnection  of magnetic field lines is required, e.g.,
MHD instabilities at the dissipation radius or large-scale changes in the 
magnetic field polarity of the flow \citep{romanova_92,spruit_01,spruit_02,lyutikov_06,giannios_10,zhang_11,mckinney_12,parfrey_15}.

To minimize the energetic requirements of the dissipation mechanism, we assume, for both shocks and reconnection, 
that relativistic electrons ($\gamma_{\rm e} \gg 1$) are responsible for the observed
radiation and that they lose most of their energy radiatively. 
In what follows, therefore, the terms radiative and dissipative efficiency will be used interchangeably.
We also assume that electrons pick up a large fraction of the dissipated energy in both models, while 
the rest of the dissipated energy goes into protons, or ions more generally. Thus, the electron energy density
$U_{\rm e}$ in the emitting region is parametrized as a fraction of the overall energy density $e$, i.e., 
$U_{\rm e}=\xi_{\rm e}\, e$, where $e$ does not include the rest mass energy density.  For the discussion that follows, we 
set $\xi_{\rm e}=0.5$ {for electron-proton flows and $\xi_{\rm e}=1$ for electron-positron flows}. 
These values are physically motivated on the basis of PIC simulations, 
for both relativistic shocks \citep{spitkovsky_08,sironi_spitkovsky_11a,sironi_13} and magnetic reconnection \citep[e.g.,][]{ss_14,melzani_14}.


\subsection{Shocks}
\label{shocks}
In the case of shocks, the downstream (emitting region) and upstream (large-scale jet) properties can be
directly connected through the jump conditions at the shock.
Let us first consider a hydrodynamic shock that corresponds to a non-magnetized jet.
For a cold upstream medium, the kinetic energy per unit rest mass in the downstream region is found to be $e/\rho c^2=\gamma_{\rm rel}-1$, where
$\gamma_{\rm rel} = 1/\sqrt{1-\beta_{\rm rel}^2}$ and $\beta_{\rm rel}=v_{\rm rel}/c$ is the relative velocity between the 
upstream and downstream regions, in units of the speed of light. 
The dissipative efficiency, which, under our assumptions,  is equal to the radiative one, is then
\eqb
f_{\rm sh}\equiv \frac{U_{\rm e}}{e+\rho c^2}=\xi_{\rm e}\left(1-\frac{1}{\gamma_{\rm rel}}\right),
\eqe
which in the ultra-relativistic regime ($\gamma_{\rm rel} \gg 1$) reduces to $f_{\rm sh} \approx \xi_{\rm e}$. {For our physically-motivated choice of $\xi_{\rm e}=0.5-1$, the above relation yields
 an efficiency in the range of the observed values.}

In general, the upstream region is magnetized, and in this case 
the efficiency drops, unless $\gamma_{\rm rel}$ increases accordingly. 
Thus, shocks formed in a flow with any appreciable magnetization must
be mildly relativistic or relativistic, in order to satisfy the efficiency requirement $f_{\rm sh}\gtrsim 0.1$ (see,
e.g., \citealt{mimicaetal09}). This will be shown with detailed numerical examples at the end of this section.
Similar to the hydrodynamic case, the downstream and upstream quantities can be related 
through the MHD jump conditions (for a treatment of the MHD jump  conditions in  relativistic shocks
see, e.g., \citealt{kennelCoroniti84, applcamenzind88}). 
In the following, we introduce the notation $q_i$, with $i=1$  for upstream and $i=2$  for downstream, 
to refer to quantities of region $i$ measured in the respective rest frame. 
The quantities to be determined and tested against the observational constraints are 
the efficiency ($f_{\rm sh}$) and the ratio of the 
magnetic to particle energy densities ($U_{\rm B,2}/U_{\rm e, 2}$) as a function of the upstream magnetization $\sigma_1$.
{From this point on, we will refer to $\sigma_1$ simply as $\sigma$, in order to use a uniform notation throughout the text.}
The efficiency is defined as 
 \eqb
 f_{\rm sh}=\frac{U_{\rm e,2}}{e_2+\rho_2 c^2 + U_{\rm B, 2}},
 \eqe
 where $U_{\rm e, 2}= \xi_{\rm e} e_2$, $U_{\rm B, 2}= B_2^2/8\pi$ and $\rho_2$ is the proper density of the downstream region. The ratio of magnetic to electron energy is
 \eqb
 \frac{U_{\rm B,2}}{U_{\rm e,2}} = \frac{\hat{\Gamma}-1}{\xi_{\rm e}}\frac{1}{\beta_{\rm p,2}},
 \eqe
 where $e_2 = P_2/(\hat{\Gamma}-1)$, $\hat{\Gamma}$ is the adiabatic index ranging from $5/3$ (non-relativistic ideal gas) to $4/3$ (ultra-relativistic gas), and 
$\beta_{\rm p, 2} \equiv P_2/U_{\rm B, 2}$ is the  plasma beta
parameter for the downstream region, while  $P_2$ is the gas pressure. 

The parameter $f_{\rm sh}$ as defined above represents the fraction of post-shock energy that is deposited
into electrons (or generally, pairs), and thus available to be radiated. Yet, the observational requirement of efficient dissipation ($f_{\rm sh}\gtrsim 0.1$) is not the only constraint imposed on the candidate dissipation mechanism.
Unless shocks can accelerate particles to energies well beyond the peak of their thermal distribution, 
dissipation at shocks is not a promising process for explaining the observed non-thermal emission from jets (see also Section~\ref{obs}).
In fact, it is well known that  Fermi acceleration cannot operate in the so-called superluminal shocks
(e.g., \citealt{kirkheavens89,begelmankirk90,sironi_spitkovsky_09,sironi_spitkovsky_11a,sironi_13}).
There,  particles moving along the magnetic field at the speed of light cannot outrun the shock, whose speed in the  pre-shock medium is $v_{1} \sim c$, for ultra-relativistic flows.
Since the Fermi process requires repeated crossings of the shock, this implies that 
 Fermi acceleration in superluminal relativistic shocks is extremely inefficient. For this reason, we also require the shock to be 
subluminal, or equivalently the condition $\beta_{\rm 1}/\cos\theta_1 <1$ should hold (see, e.g., eq.~(1) in \citealt{kirkheavens89}), where $\beta_1=v_1/c$  and $\theta_1$ is the angle between the shock normal and the magnetic field as measured in the upstream frame.

We solve the MHD jump conditions for a cold upstream medium\footnote{We numerically solve eqs.~(25), (28), (29), (31), (33) and (37)
 in \cite{applcamenzind88}, while we use eqs.~(13)-(14) in \citet{service86} for the dependence of the adiabatic index on the downstream temperature.} 
and two representative shock obliquities, i.e. $\theta_1=30^{\rm o}$ and $60^{\rm o}$.
The results are presented in Fig.~\ref{fig1} and our main conclusions can be summarized as follows:

\begin{itemize}
\item For $\sigma \lesssim 0.1$, the efficiency does not strongly depend on the flow magnetization. 
\item The shock must be mildly or ultra-relativistic in order to be efficient,
  since $f_{\rm sh}\gtrsim 0.1$ only for $\gamma_{\rm rel} \beta_{\rm rel}
  \gtrsim 1$, which is consistent with the argument by \citet{rees_94}.
\item The ratio of magnetic to particle energy densities {in the downstream} increases
 {with the upstream magnetization}. 
For $\sigma \simeq 0.001-0.1$, there is a small range in $\gamma_{\rm rel} \beta_{\rm rel}$ (dependent on $\sigma$) leading to 
$0.3 \le U_{\rm B, 2}/U_{\rm e, 2} \le 3$.  This conclusion does not depend 
significantly on the magnetic field inclination relative to the shock front.
\item For oblique shocks with angles $\theta_1 \lesssim 30^{\rm o}$, there is a small parameter space
where all of the three constraints are satisfied, namely $0.03 \le f_{\rm sh} \le 0.3$, $0.3 \le U_{\rm B,2}/U_{\rm e,2}\le 3$ and the shock is subluminal.
This parameter regime involves mildly relativistic shocks with $\beta_{\rm rel}\gamma_{\rm rel}\simeq 1$ and
 moderate magnetization $\sigma \sim 0.1$.
 \item For substantial magnetic field inclination ($\theta_1 > 45^{\rm o}$), there
is a strong tension between the efficiency constraint,
that favors fast shocks, and the subluminality requirement. 
For $\theta_1 =60^{\rm o}$, we find that the efficiency for subluminal
shocks is at most $\sim 3\%$.
\end{itemize}
\begin{figure}
 \centering
 \includegraphics[width=0.48\textwidth]{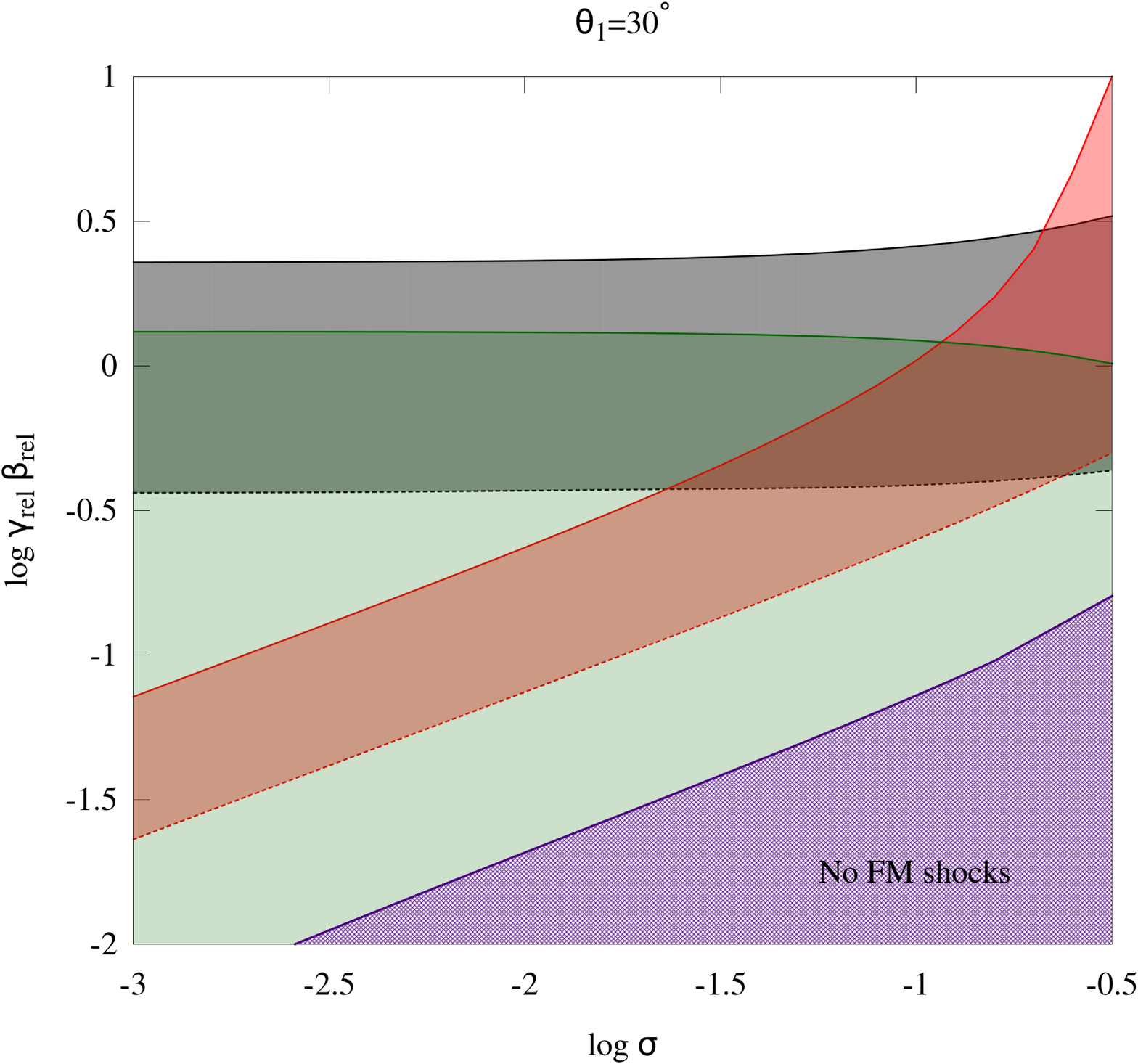}\\
  \includegraphics[width=0.48\textwidth]{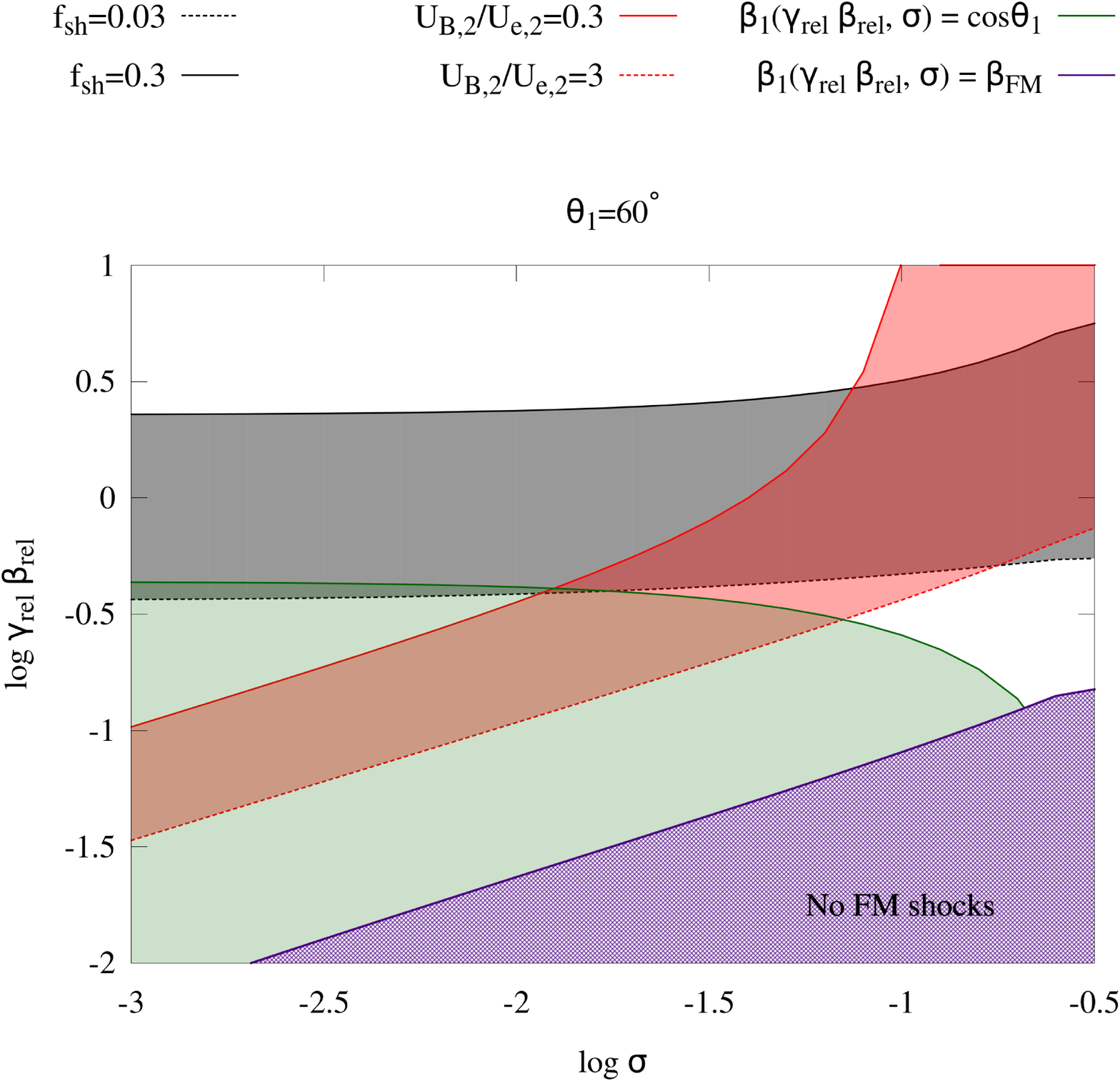}
 \caption{Plot of $\gamma_{\rm rel} \beta_{\rm rel}$ as a function of the upstream magnetization $\sigma$ for two
different inclinations of the upstream magnetic field relative  to the shock normal (measured in the upstream frame): $\theta_1=30^{\rm o}$  (top panel) 
and $60^{\rm o}$ (bottom panel). The black, red and green colored regions lead, respectively,
to efficiencies 
$0.03 \le f_{\rm sh} \le 0.3$, ratios of magnetic to electron energy densities $0.3 \le U_{\rm B,2}/U_{\rm e,2} \le 3$ 
and subluminal configurations. For parameters lying
in the purple-hatched region no fast magnetosonic shock solutions can be found. In both panels we assume 
$\xi_{\rm e}=0.5$, as appropriate for electron-proton flows \citep{sironi_spitkovsky_11a,sironi_13}. }
 \label{fig1}
\end{figure}

The analysis above assumes that the magnetic field in the downstream  region
is merely the result of shock compression (as opposed to fields generated
via plasma instabilities at the shock front) and that the upstream region is 
sufficiently magnetized for the 
superluminal constraint to hold, i.e., $\sigma\gtrsim 10^{-3}$ in electron-positron shocks and $\sigma\gtrsim 10^{-4}$
in electron-proton shocks \citep{sironi_13}.
{We showed that the observational requirements of  $0.03 \lesssim f_{\rm sh} \lesssim 0.3$, 
$0.3 \lesssim U_{\rm B,2}/U_{\rm e,2}\lesssim 3$ and the subluminality constraint for efficient particle acceleration 
are simultaneously satisfied only for certain shock obliquities and for $\sigma >0.01$.} In this case, 
the downstream field is likely to be dominated by the shock-compressed component, in agreement with our assumptions.

These assumptions, however, may not hold in other environments, such as the external shocks of GRBs. There,
the shock propagates into the interstellar medium, whose magnetization is extremely small ($\sigma \sim 10^{-9}$).
 Modelling of the GRB afterglow emission implies that the emitting region is far from equipartition, having 
$U_{\rm e}\gg U_{\rm B}$ \citep{kumar_09,kumar_10,lemoine_13,sironi_13}. Thus, in the context of GRBs, our arguments are applicable 
only to the ``internal'' jet emission  mechanism and {not to the afterglow phase.}
    

\begin{figure*}
 \centering
  \includegraphics[width=\textwidth]{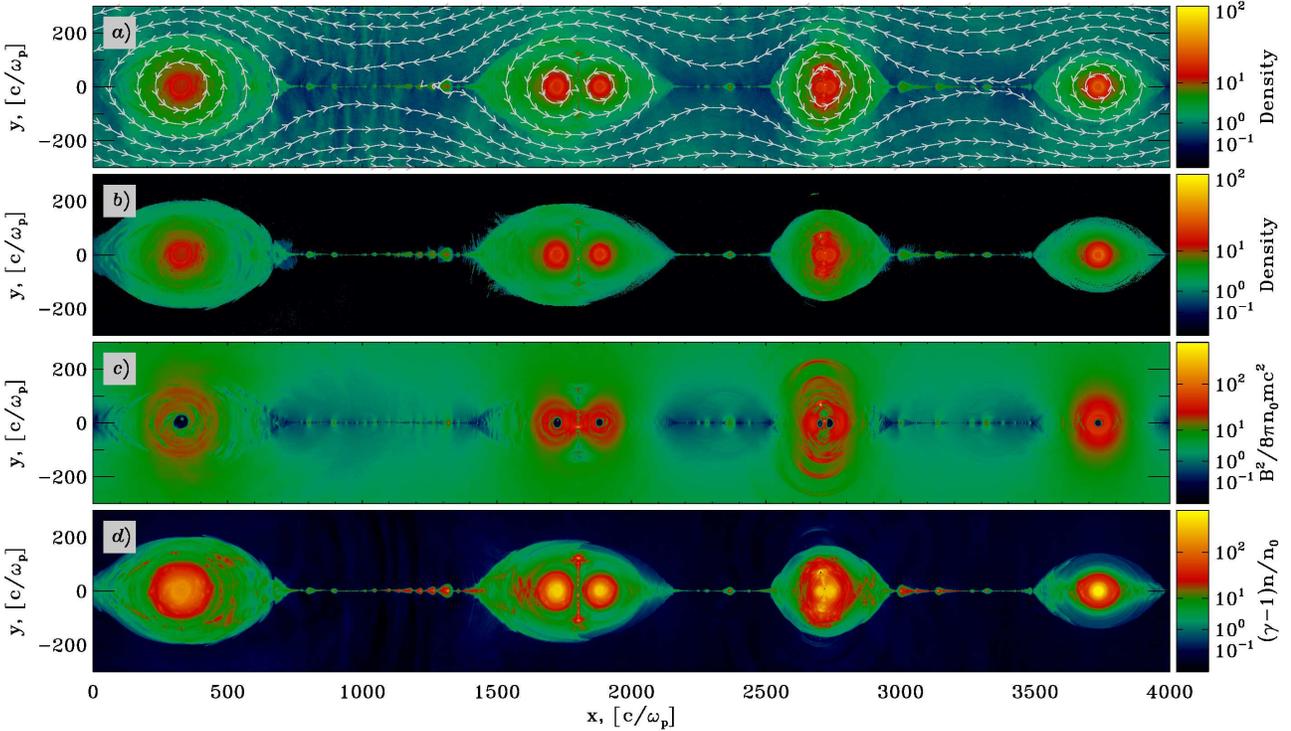}
 \caption{Structure of the reconnection layer at $\omega_{\rm p} t=3000$, from a 2D simulation of $\sigma=10$ anti-parallel 
 reconnection in electron-positron plasmas, as described in \citet{ss_14}. 
 The box extends along $x$ over $\sim 6550\comp$ (65536 cells), and along $y$ over $\sim 6000\comp$ ($\sim60000$ cells), but we only show a subset of the domain, 
 to emphasize the small-scale structures in the reconnection layer. Here, $\comp=\sqrt{m c^2/4 \pi n_0 e^2}$ is the plasma skin depth, where $n_0$ is the particle number density far from the reconnection layer. We present (a) 
 the particle number density in units of $n_0$, with overplotted magnetic field lines; 
 (b) the particle density, with artificially colored in black the regions that do not meet the selection criterion 
 $\gamma>1.1$ for the reconnection downstream; (c) the magnetic energy density normalized to the rest mass energy density far from 
 the reconnection layer, $\epsilon_B=B^2/8\pi \rho c^2$, where $\rho = m \,n_0$ is the mass
 density far from the reconnection layer; and (d) the kinetic energy density normalized to the rest mass energy density far from the reconnection layer.}
 \label{fig:fig1}
\end{figure*}

\begin{figure}
\includegraphics[scale=0.6]{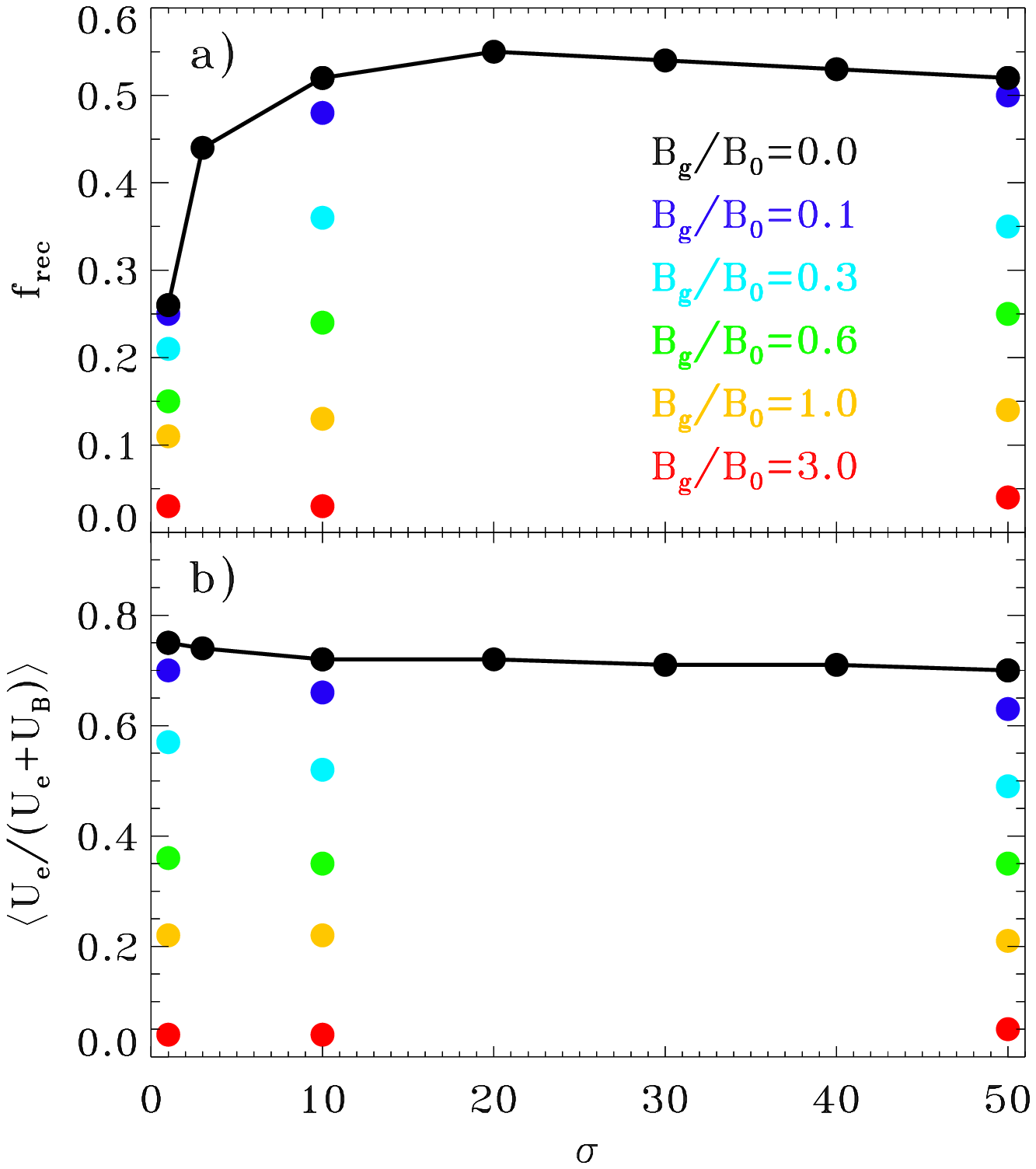}
\caption{As a function of the flow magnetization, 
we plot the dissipative efficiency (top panel) and the volume-averaged 
$\langle U_{\rm e}/(U_{\rm e}+U_{\rm B})\rangle$ (bottom panel) for 
relativistic reconnection in electron-positron plasmas, from a series of 
2D simulations as described in \citet{ss_14}. The black lines refer to the case 
of anti-parallel fields (so, $B_{\rm g}/B_0=0$, where $B_{\rm g}$ is the guide field and $B_0$ is the alternating field), 
whereas the effect of the guide field is shown by the filled colored circles, as indicated in the legend. {
We note that the range of values
$0.3 \lesssim {U_{\rm e}/U_{\rm B}}\lesssim 3$ inferred from the observations corresponds to $0.2\lesssim U_{\rm e}/(U_{\rm e}+U_{\rm B})\lesssim 0.75$
for the case of homogeneous plasmoids.}}
\label{fig:rec1}
\end{figure}

\subsection{Magnetic reconnection}
In reconnection, magnetic field energy is transformed into particle energy.
Assuming that half of the magnetic energy is converted into internal energy (quite a realistic assumption, as we demonstrate below with PIC simulations), and assuming fast cooling
electrons, the expected radiative efficiency is $f_{\rm rec}\sim 0.5\,\xi_{\rm e}\,\sigma/(\sigma+2)$.
{For $\xi_{\rm e}\sim  0.5-1$, as found in PIC simulations of electron-positron (yielding $\xi_{\rm e}\sim 1$) and electron-proton (giving $\xi_{\rm e}\sim 0.5$) reconnection  \citep[e.g.,][]{ss_14,melzani_14}, this implies that $\sigma \gtrsim 1$ for}
reconnection to be energetically viable in powering the jet emission.
Hence, we deal with {\it relativistic reconnection}, 
in which the mean magnetic energy per particle is larger than the rest mass energy, or equivalently $\sigma\gtrsim 1$. 

Recent progress in the study of relativistic reconnection via PIC simulations has been 
remarkable \citep[see][for a review]{kagan_15}, demonstrating that reconnection in high-$\sigma$ plasma is fast and unsteady. The reconnection downstream is composed of plasmoids that contain
the reconnected plasma. In two dimensions (2D), 
they appear as overdense magnetic islands (see the 2D density pattern in the top panel of \fig{fig1}) connected by thin X-lines. 
The plasmoid instability further fragments each X-line into a series of smaller islands, separated 
by X-points (e.g., see the small islands at $700\, c/\omega_{\rm p}\lesssim x\lesssim 1400\, c/\omega_{\rm p}$ in 
the top panel of \fig{fig1}, where $c/\omega_{\rm p}$ is the electron skin depth). At the X-points, the particles are not tied to the magnetic field lines  
and they get accelerated along the reconnection electric field. 
The late-time particle spectrum integrated over the whole reconnection region is a power-law, whose 
slope is $-2$ for $\sigma=10$ (as  commonly 
assumed in blazar SED modeling), and becomes harder with increasing magnetization (\citealt{ss_14}; 
see also, e.g., \citealt{zenitani_01,jaroschek_04,bessho_12,kagan_13,cerutti_13b,guo_14,werner_14}). 
Efficient particle acceleration to non-thermal energies is a generic by-product of 
the long-term evolution of relativistic reconnection in both two and three dimensions. 
In three dimensions (3D), the so-called drift-kink mode corrugates the reconnection layer at 
early times \citep{daughton_98,zenitani_08}, but the long-term evolution is controlled by the plasmoid instability, 
that facilitates efficient particle acceleration, in analogy to the two-dimensional physics \citep{ss_14}.

In summary, recent PIC simulations of relativistic 
reconnection have convincingly demonstrated that for $\sigma\gtrsim 1$ 
(of interest for relativistic jets), the accelerated particles populate
extended non-thermal power-laws. We now investigate 
whether relativistic reconnection can provide the required efficiency 
$f_{\rm rec}$ and kinetic-to-magnetic energy ratio $U_{\rm e}/U_{\rm B}$ inferred from jet observations.


The dissipated magnetic energy in the reconnection downstream is 
distributed between particles and magnetic fields. The {plasmoids appear as 
inhomogeneous structures}, with the core dominated by particle energy (panel (d) in \fig{fig1}), 
whereas the outskirts are magnetically dominated (panel (c) in \fig{fig1}).
{Because of the inhomogeneity of the plasmoids, it is meaningful to define
the dissipative efficiency 
as a volume- or surface-averaged quantity in 3D and 2D, respectively, summed over many plasmoids:} 
\be\label{eq:rec1}
f_{\rm rec}\equiv \frac{\sum_{i}\int_{V_{i}} U_{\rm e}{\rm d}V_{ i}}{\sum_{i}\int_{V_{i}} (e+\rho c^2+U_{\rm B})\,{\rm d}V_{ i}}
\ee
where $V_{i}$ is the volume (or surface) of each plasmoid, $U_{\rm e}$ is the kinetic energy of the emitting leptons 
(electrons alone for  electron-proton reconnection, or both species for electron-positron reconnection), {while the sum runs
over all the plasmoids in the reconnection layer (see the colored regions in \fig{fig1}(b)).

{The fact that the reconnection upstream is taken to be cold
facilitates the identification of the regions where dissipation of
energy has taken place, as we now explain.}
We select the plasmoid volume $V_i$  where to compute the
integrals above (or actually surface, for 2D simulations as in \fig{fig1}) by selecting all the regions where the mean particle
 Lorentz factor is $\gamma>1.1$ (more precisely, in electron-ion reconnection we employ here the mean {\it ion} Lorentz factor). The inflow region is cold and moves at the reconnection speed, 
 which for $\sigma\gtrsim 10$ is $v_{\rm rec}\sim 0.15\,c$ \citep[e.g.,][]{ss_14}, so that the mean Lorentz
 factor there is only $\gamma_{\rm rec}\sim 1.01$, which is below the threshold we employ. In \fig{fig1}(b), we only display the regions
 where the condition $\gamma>1.1$ is met (otherwise, the color is artificially set
 as black), showing that this criterion provides an excellent
 identification of the reconnection downstream (i.e., of the magnetic
 islands). We have employed this criterion across the whole range of
 magnetizations we have investigated ($\sigma=1-50$) and for different
 strengths of the guide field  $B_{\rm g}$ orthogonal to the 
annihilating fields  $B_0 $ (from $B_{\rm g}/B_0=0$ up to 3), 
finding always an excellent spatial correlation with the reconnection downstream. In the following, we define $\sigma=B_0^2/4\pi \rho c^2$, i.e., excluding the contribution of the guide field.


We now explain how to extract the equipartition parameter $U_{\rm e}/U_{\rm B}$. Fits to observations {are performed under the assumption of} a homogeneous emitting region 
({the so-called one-zone homogeneous emission models). They} infer
$0.3 \lesssim {U_{\rm e}/U_{\rm B}}\lesssim 3$, which can also be expressed as
$0.2\lesssim U_{\rm e}/(U_{\rm e}+U_{\rm B})\lesssim 0.75$. 
Since the reconnection downstream is inhomogeneous (see the magnetic islands in \fig{fig1}),
we need to define a measure of $U_{\rm e}/(U_{\rm e}+U_{\rm B})$ in the downstream region that closely corresponds to what the homogeneous models probe.
{In the limit of fast cooling particles, 
we should calculate the volume-average (surface-average in 2D simulations) 
of $U_{\rm e}/(U_{\rm e}+U_{\rm B})$ weighted with the electron energy density $U_{\rm e}$, and then sum over
all the plasmoids:}
\be\label{eq:rec2}
\left\langle \frac{U_{\rm e}}{U_{\rm e}+U_{\rm B}} 
\right\rangle\equiv\frac{\sum_{i} \int_{V_{i}} U_{\rm e}  \frac{U_{\rm e}}{U_{\rm e}+U_{\rm B}}{\rm d}V_{i}}{\sum_{i}\int_{V_{i}} U_{\rm e} {\rm d}V_{i}}.
\ee
The reason for our choice of the weighting factor $U_{\rm e}$ is that regions 
with low electron energy density cannot contribute much to the emission. Although
in regions where $U_{\rm B}$ is larger than $U_{\rm e}$ (e.g. in the outskirts 
of the plasmoids, compare \fig{fig1}(c) and (d)), the synchrotron cooling rate
is significantly higher than in regions with lower $U_{\rm B}$ (e.g.,
the plasmoid cores), the total radiated power in the fast cooling regime
is always limited by the one injected into the
emitting particles.} As a side note, we remark that the choice to integrate $U_{\rm e}/(U_{\rm e}+U_{\rm B})$ rather than $U_{\rm e}/U_{\rm B}$ in \eq{rec2} is motivated by the fact that the integral would otherwise be largely dominated by the very center of magnetic islands, where $U_{\rm B}\ll U_{\rm e}$ (see \fig{fig1}(c) and (d)). Our choice  is then appropriate for comparing to observational inferences based on homogeneous emission models.

  
      \begin{table}
\begin{center}
\vspace{0.05 in}\caption{Results from 2D PIC simulations in electron-positron plasmas without guide field, as a function of the flow magnetization.}
\label{table:sigma1}
\begin{tabular}{cccccccc}
\hline \hline
\multicolumn{1}{c}{$\sigma$} &
\multicolumn{1}{c}{1} & 
\multicolumn{1}{c}{3} &
\multicolumn{1}{c}{10} &
\multicolumn{1}{c}{20} &
\multicolumn{1}{c}{30} &
\multicolumn{1}{c}{40} &
\multicolumn{1}{c}{50}
\\
\hline
\vspace{2pt}
$f_{\rm rec}$ & 0.26 & 0.44 & 0.52 & 0.55 & 0.54 & 0.53 & 0.52\\ 
$\langle U_{\rm e}/(U_{\rm e}+U_{\rm B})\rangle $ & 0.75 & 0.74 & 0.72& 0.72& 0.71& 0.71 & 0.70\\ 
\hline
\end{tabular}
\end{center}
\end{table}

\begin{table}
\begin{center}
\vspace{0.05 in}\caption{Results from 2D PIC simulations in electron-positron plasmas, as a function of the flow magnetization and the guide field strength.}
\label{table:guide}
\begin{tabular}{lccccccc}
\hline \hline
\multicolumn{1}{c}{$$} &
\multicolumn{1}{c}{$B_{\rm g}/B_{0}$} &
\multicolumn{1}{c}{0} & 
\multicolumn{1}{c}{0.1} &
\multicolumn{1}{c}{0.3} &
\multicolumn{1}{c}{0.6} &
\multicolumn{1}{c}{1} &
\multicolumn{1}{c}{3}
\\
\hline
\vspace{2pt}
$\!\!\!\!\sigma=1$ & $f_{\rm rec}$ & 0.26 & 0.25 & 0.21 & 0.15 & 0.11 & 0.03\\ 
$\!\!\!\!\sigma=1$ & $\!\!\langle U_{\rm e}/(U_{\rm e}+U_{\rm B}) \rangle\!\!$ & 0.75 & 0.70 & 0.57 & 0.36 & 0.22 & 0.04\\ 
\hline
\vspace{2pt}
$\!\!\!\!\sigma=10$ & $f_{\rm rec}$ & 0.52 & 0.48 & 0.36 & 0.24 & 0.13 & 0.03\\ 
$\!\!\!\!\sigma=10$ & $\!\!\langle U_{\rm e}/(U_{\rm e}+U_{\rm B}) \rangle\!\!$ & 0.72 & 0.66 & 0.52 & 0.35 & 0.22 & 0.04\\ 
\hline
\vspace{2pt}
$\!\!\!\!\sigma=50$ & $f_{\rm rec}$ & 0.52 & 0.50 & 0.35 & 0.25 & 0.14 & 0.04\\ 
$\!\!\!\!\sigma=50$ & $\!\!\langle U_{\rm e}/(U_{\rm e}+U_{\rm B}) \rangle\!\!$ & 0.70 & 0.63 & 0.49 & 0.35 & 0.21 & 0.05\\ 
\hline
\end{tabular}
\end{center}
\end{table}

By integrating over the surface of the magnetic islands, we compute $f_{\rm rec}$ and $\langle U_{\rm e}/(U_{\rm e}+U_{\rm B}) \rangle$ as described above.  Our main results are shown in \fig{rec1} for the case of electron-positron reconnection and in \fig{rec2} for electron-proton reconnection. Our conclusions can be summarized as follows (see also Tables \ref{table:sigma1}-\ref{table:sigma2}):
\begin{itemize}

\item Jet observations do not impose strong constrains on the
  reconnection model {as long as the reconnection is relativistic}.
For $\sigma \gtrsim 1$ and guide fields weaker that the reconnecting
field ($B_{\rm g}\lesssim B_0$), all constraints are satisfied. We have checked that this conclusion holds
for a broad range of conditions: pair and electron-ion plasma, and upstream 
magnetizations as large as $\sigma \sim 50$. 

\item In electron-positron plasmas (top panel in \fig{rec1}), the reconnection efficiency 
asymptotically approaches $f_{\rm rec}\sim 0.5$ for $\sigma\gtrsim 10$ {and no guide field.} In electron-ion plasmas 
(colored lines in the top panel of \fig{rec2}), it is roughly half of that value, since in the numerator of \eq{rec1} 
only electrons contribute (as opposed to both species, for electron-positron plasmas). For
lower $\sigma$, the rest mass energy appreciably contributes   
to the denominator of \eq{rec1}, which results in a decrease of the reconnection efficiency.

\item As a result of the efficient field dissipation, the energy density of the radiating particles 
is {\it larger} than that of the magnetic field at the emitting region
by a factor of a few, both in electron-positron plasmas (black line in the bottom panel of \fig{rec1}) 
and electron-ion plasmas (colored lines in the bottom panel of \fig{rec2}). Once again, the difference 
between the colored lines and the black line in \fig{rec2} is due to the fact that, in electron-positron 
reconnection (black line), both species contribute to the kinetic energy $U_{\rm e}$ of emitting particles that enters in \eq{rec2}.



\item For weak guide fields ($B_{\rm g}/B_0 \lesssim 0.5$), the magnetic energy is very efficiently 
dissipated ($f_{\rm rec} \gtrsim 0.2$ in electron-positron plasmas).  For progressively stronger guide fields, the radiative 
efficiency drops (see the colored circles in the top panel of \fig{rec1}) and the
emitting regions turn magnetically dominated (colored circles in the bottom panel). 
The transition happens gradually around 
$B_{\rm g}/B_{\rm 0}\sim 0.5$, independently of $\sigma$. 

\item For electron-positron reconnection, we find that the ratio between particle kinetic energy and magnetic energy has an upper limit of $U_{\rm e}/U_{\rm B}\sim 3$, see \fig{rec1}(b) (and $U_{\rm e}/U_{\rm B}\sim 2.5$  for electron-proton plasmas, see \fig{rec2}(b)). This is independent of $\sigma$ and is reached in the limit of weak guide fields. When increasing the guide field strength, the emission region gets more magnetized and $U_{\rm e}/U_{\rm B}$ can become arbitrarily small. However, we remark that when the emission region is strongly magnetized, the dissipation process is inefficient (i.e., $f_{\rm rec}\ll1$), so the resulting observational signature is very weak. It follows that in reconnection-dominated systems  there is an observational bias to infer equipartition. 

\item We can provide a convenient parameterization of the results presented in \fig{rec1} and \fig{rec2} (see also Tables \ref{table:sigma1}-\ref{table:sigma2}), assuming that a fraction $\sim 55\%$ of the magnetic energy in alternating fields (so, excluding the guide field) is converted into particle kinetic energy, and that the guide field energy is not dissipated. We obtain 
\be
f_{\rm rec}=0.55\, \xi_{\rm e}\frac{\sigma}{\sigma\,[1+(B_{\rm g}/B_0)^2]+2}
\ee
where $\xi_{\rm e}\sim1$ in electron-positron plasmas and $\xi_{\rm e}\sim0.5$ in electron-proton plasmas. Similarly, the expected equipartition parameter is
\be
 \frac{U_{\rm e}}{U_{\rm e}+U_{\rm B}} =1.3\, \frac{0.55\, \xi_{\rm e}}{0.55\,\xi_{\rm e}+0.45+(B_{\rm g}/B_0)^2}
\ee
where the extra factor of 1.3 accounts for the fact that the integrand in \eq{rec2} has been weighted with the electron energy density $U_{\rm e}$. These two formulae provide a satisfactory description of the trends observed in \fig{rec1} and \fig{rec2}.

\item Our results for electron-ion reconnection 
are nearly independent of the numerical choice of the mass ratio, which is 
constrained in PIC simulations to be smaller than the realistic value, for computational
convenience. In fact, the three colored lines in \fig{rec2} (blue for $m_{\rm p}/m_{\rm e}=6.25$, green 
for $m_{\rm p}/m_{\rm e}=25$ and red for $m_{\rm p}/m_{\rm e}=100$) nearly overlap. It follows that our results 
can be confidently applied to the case of a realistic mass ratio.
\end{itemize}

\begin{figure}
\includegraphics[scale=0.6]{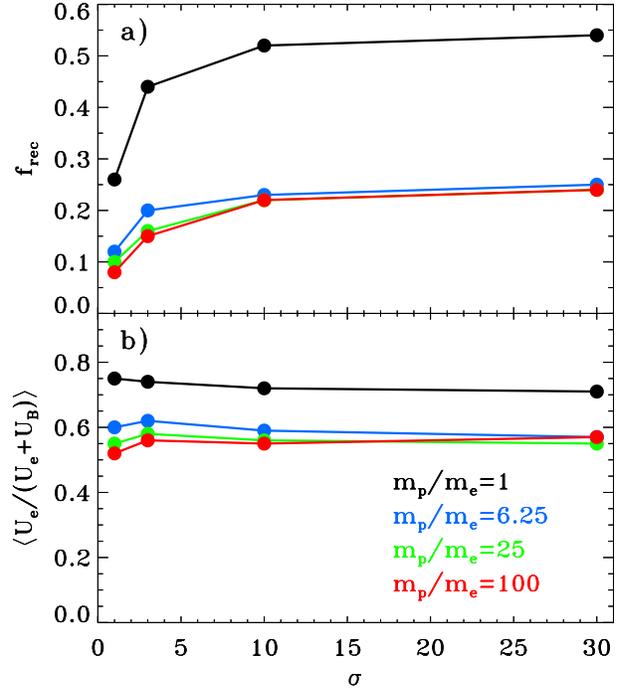}
\caption{As a function of the flow magnetization, we plot the dissipative efficiency (top panel) and the volume-averaged $\langle U_{\rm e}/(U_{\rm e}+U_{\rm B})\rangle$ (bottom panel) for relativistic reconnection in electron-ion plasmas, from a series of 2D simulations as described in \citet{ss_14}. The black lines refer to the case of electron-positron reconnection, for comparison. The dependence on the numerical choice of the mass ratio $m_{\rm p}/m_{\rm e}$ is illustrated by the colored curves, as indicated in the legend.}
\label{fig:rec2}
\end{figure}

\begin{table}
\begin{center}
\vspace{0.05 in}\caption{Results from 2D PIC simulations in electron-ion plasmas ($m_{\rm p}/m_{\rm e}=100$) without guide field, as a function of the flow magnetization.}
\label{table:sigma2}
\begin{tabular}{ccccc}
\hline \hline
\multicolumn{1}{c}{$\sigma$} &
\multicolumn{1}{c}{1} & 
\multicolumn{1}{c}{3} &
\multicolumn{1}{c}{10} &
\multicolumn{1}{c}{30}
\\
\hline
\vspace{2pt}
$f_{\rm rec}$ & 0.08 & 0.15 & 0.22 & 0.24 \\ 
$\langle U_{\rm e}/(U_{\rm e}+U_{\rm B})\rangle $ & 0.52 & 0.56 & 0.55& 0.57\\ 
\hline
\end{tabular}
\end{center}
\end{table}

\section{Discussion and Conclusions}
\label{conclusions}

For a long time, theoretical modeling of relativistic astrophysical jets has faced a conundrum:
if jets are launched as magnetically-dominated flows, as most theories predict, why do they appear
to be in rough equipartition between magnetic fields and radiating particles at 
the emission region? Does the Poynting flux convert into bulk kinetic energy prior to the jet emission,
and then shock waves are responsible for accelerating the radiating particles? 

The answer is: probably not. For shocks to be efficient in powering the jet emission, we find that they need to be
at least mildly relativistic, i.e.,  $\gamma_{\rm rel}\gtrsim $ a few. To attain rough equipartition between $U_{\rm e}$ and $U_{\rm B}$ in the shock downstream, as inferred from modeling the SED of blazar sources,
the magnetization in the shock upstream has to be in the range $\sigma\sim 0.01-0.1$.
Under these conditions, for most magnetic field orientations, shocks are superluminal.
Charged particles are constrained to follow the field lines, whose orientation prohibits repeated crossings of the shock. So, the particles have no chance to undergo Fermi acceleration, which instead is required to explain the broadband non-thermal emission signatures of blazar jets. We find that the shock model can work only in a small region
of the parameter space, with $\gamma_{\rm rel}\sim 1.5$ and $\theta_1\lesssim 30^\circ$.

If the jet remains magnetically-dominated at the dissipation distance (as most models of MHD jets predict), dissipation
through reconnection is more likely. PIC simulations of magnetic reconnection in high-$\sigma$ plasmas (see \citealt{kagan_15}, for a review) can now describe the evolution
of the system for long enough times to probe the dynamics of the downstream region as well
as the physics of particle acceleration. The simulations have demonstrated that high-$\sigma$ reconnection 
has several attractive features to explain the observations. 
The reconnection process is fairly fast, because of the 
fragmentation of the layer into plasmoid chains. The plasmoids (or magnetic islands) contain the reconnection fluid
and qualify as the emitting regions in this model. 
For $\sigma\gtrsim 1$, the particles are efficiently accelerated in the course of the reconnection process \citep[e.g.,][]{ss_14}.  In agreement with the observations, the particle energy spectrum is a power-law non-thermal distribution, with a harder power-law slope for higher $\sigma$.
In this work, we have shown that the reconnection process is particularly efficient in depositing magnetic energy 
into non-thermal particles, and at the same time that the emitting regions are in rough equipartition
between particles and magnetic field.

From a suite of 2D PIC simulations of relativistic reconnection, we have found that,
for $\sigma\gtrsim 1$ and zero guide field, the emitting regions (plasmoids)
have $U_{\rm e}/U_{\rm B}\sim 3$ and that a significant fraction of the the upstream energy is deposited
into energetic electrons. More precisely, the dissipative efficiency reaches $f_{\rm rec}\sim 0.5$ for electron-positron plasmas (where both species can contribute to the emission)  
and $f_{\rm rec}\sim 0.2$ in electron-proton plasmas. 
In general, the presence of a strong guide field makes the process slower and less efficient, and the reconnection downstream becomes more magnetically-dominated.
Nevertheless, for a broad range of guide field strengths $B_{\rm g}/B_{0}\lesssim 0.5$,
we find that $U_{\rm e}/U_{\rm B}\gtrsim 0.3$ and $f_{\rm rec}\gtrsim 0.1$, which is within the range inferred from the observations.\footnote{We point out that our reconnection model predicts an upper limit for the ratio of electron energy to magnetic energy of $U_{\rm e}/U_{\rm B}\sim 3$. It follows that, in its simplest formulation, it cannot account for the few sources where $U_{\rm e}/U_{\rm B}\gg1$ \citep[e.g.,][]{celotti_08,ghisellini_14}.}
This conclusions holds regardless of the flow magnetization, in the regime $\sigma \gtrsim 1$ of relativistic reconnection (we tested from $\sigma=1$ up to
$\sigma=50$). In fact, the measured ratio $U_{\rm e}/U_{\rm B}$ might be used to
directly probe the strength of the guide field in the reconnection region (see Figs.~\fign{rec1} and \fign{rec2}).
 
The fact that $\sigma\gtrsim 1$ also implies that 
the emitting regions may be characterized by relativistic motions, with outflow Lorentz factors $\Gamma_{\rm out}\sim \sqrt{\sigma}$
as measured in the rest frame of the jet \citep{lyubarsky_05}, which has profound implications for jet variability
\citep[e.g.,][]{lyutikov_03,giannios_09,nalewajko_11,narayan_12,giannios_13}.
 
In summary, we have shown that shocks are unlikely to mediate the dissipation of energy in relativistic jets, mainly because the efficiency and subluminality constraints cannot be satisfied simultaneously. If the bulk of the jet has $\sigma \gtrsim 1$ and 
sufficiently small-scale fields prone to be dissipated, it appears that magnetic reconnection can
 satisfy all the basic conditions for the emission: extended particle distributions,
efficient dissipation and rough equipartition between particles and magnetic field 
in the emitting region. Dissipation via reconnection is a promissing process to explain the multi-wavelength non-thermal emission of relativistic jets.


\section*{Acknowledgments}
L.S. is supported by NASA through Einstein
Postdoctoral Fellowship grant number PF1-120090 awarded by the Chandra
X-ray Center, which is operated by the Smithsonian Astrophysical
Observatory for NASA under contract NAS8-03060. M.P. is supported by
NASA through Einstein Postdoctoral Fellowship grant number PF3-140113
awarded by the Chandra X-ray Center, which is operated by the
Smithsonian Astrophysical Observatory for NASA under contract
NAS8-03060. DG acknowledges support from NASA grant NNX13AP13G. 
The simulations were  performed on XSEDE resources under
contract No. TG-AST120010, and on NASA High-End Computing (HEC)
resources through the NASA Advanced Supercomputing (NAS) Division at 
Ames Research Center. 

\bibliography{reco_vs_shock_ref}

\begin{thebibliography}{}

\bibitem[\protect\citeauthoryear{{Abdo}, {Ackermann}, {Ajello}, {Baldini},
  {Ballet}, {Barbiellini}, {Bastieri}, {Bechtol}, {Bellazzini}, {Berenji} \& et
  al.}{{Abdo} et~al.}{2011}]{abdo_11}
{Abdo} A.~A.,  {Ackermann} M.,  {Ajello} M.,  {Baldini} L.,  {Ballet} J.,
  {Barbiellini} G.,  {Bastieri} D.,  {Bechtol} K.,  {Bellazzini} R.,  {Berenji}
  B.,    et al. 2011, \apj, 736, 131

\bibitem[\protect\citeauthoryear{{Aharonian}}{{Aharonian}}{2000}]{aharonian_00}
{Aharonian} F.~A.,  2000, New Astron., 5, 377

\bibitem[\protect\citeauthoryear{{Aleksi{\'c}}}{{Aleksi{\'c}}}{2012}]{aleksic_12}
{Aleksi{\'c}} J. e.~a.,  2012, \apj, 748, 46

\bibitem[\protect\citeauthoryear{{Appl} \& {Camenzind}}{{Appl} \&
  {Camenzind}}{1988}]{applcamenzind88}
{Appl} S.,  {Camenzind} M.,  1988, \aap, 206, 258

\bibitem[\protect\citeauthoryear{{Begelman} \& {Kirk}}{{Begelman} \&
  {Kirk}}{1990}]{begelmankirk90}
{Begelman} M.~C.,  {Kirk} J.~G.,  1990, \apj, 353, 66

\bibitem[\protect\citeauthoryear{{Berger}, {Sari}, {Frail}, {Kulkarni},
  {Bertoldi}, {Peck}, {Menten}, {Shepherd}, {Moriarty-Schieven}, {Pooley},
  {Bloom}, {Diercks}, {Galama} \& {Hurley}}{{Berger} et~al.}{2000}]{berger00}
{Berger} E.,  {Sari} R.,  {Frail} D.~A.,  {Kulkarni} S.~R.,  {Bertoldi} F.,
  {Peck} A.~B.,  {Menten} K.~M.,  {Shepherd} D.~S.,  {Moriarty-Schieven} G.~H.,
   {Pooley} G.,  {Bloom} J.~S.,  {Diercks} A.,  {Galama} T.~J.,    {Hurley} K.,
   2000, \apj, 545, 56

\bibitem[\protect\citeauthoryear{{Bessho} \& {Bhattacharjee}}{{Bessho} \&
  {Bhattacharjee}}{2012}]{bessho_12}
{Bessho} N.,  {Bhattacharjee} A.,  2012, \apj, 750, 129

\bibitem[\protect\citeauthoryear{{Blandford} \& {Payne}}{{Blandford} \&
  {Payne}}{1982}]{blandford_82}
{Blandford} R.~D.,  {Payne} D.~G.,  1982, \mnras, 199, 883

\bibitem[\protect\citeauthoryear{{Blandford} \& {Znajek}}{{Blandford} \&
  {Znajek}}{1977}]{blandford_77}
{Blandford} R.~D.,  {Znajek} R.~L.,  1977, \mnras, 179, 433

\bibitem[\protect\citeauthoryear{{B{\"o}ttcher}, {Reimer}, {Sweeney} \&
  {Prakash}}{{B{\"o}ttcher} et~al.}{2013}]{boettcher_13}
{B{\"o}ttcher} M.,  {Reimer} A.,  {Sweeney} K.,    {Prakash} A.,  2013, \apj,
  768, 54

\bibitem[\protect\citeauthoryear{{Celotti} \& {Ghisellini}}{{Celotti} \&
  {Ghisellini}}{2008}]{celotti_08}
{Celotti} A.,  {Ghisellini} G.,  2008, \mnras, 385, 283

\bibitem[\protect\citeauthoryear{{Cerutti}, {Werner}, {Uzdensky} \&
  {Begelman}}{{Cerutti} et~al.}{2014}]{cerutti_13b}
{Cerutti} B.,  {Werner} G.~R.,  {Uzdensky} D.~A.,    {Begelman} M.~C.,  2014,
  \apj, 782, 104

\bibitem[\protect\citeauthoryear{{Coppi} \& {Aharonian}}{{Coppi} \&
  {Aharonian}}{1999}]{coppi_99}
{Coppi} P.~S.,  {Aharonian} F.~A.,  1999, \apjl, 521, L33

\bibitem[\protect\citeauthoryear{{Costamante}, {Ghisellini}, {Giommi},
  {Tagliaferri}, {Celotti}, {Chiaberge}, {Fossati}, {Maraschi}, {Tavecchio},
  {Treves} \& {Wolter}}{{Costamante} et~al.}{2001}]{costamante_01}
{Costamante} L.,  {Ghisellini} G.,  {Giommi} P.,  {Tagliaferri} G.,  {Celotti}
  A.,  {Chiaberge} M.,  {Fossati} G.,  {Maraschi} L.,  {Tavecchio} F.,
  {Treves} A.,    {Wolter} A.,  2001, \aap, 371, 512

\bibitem[\protect\citeauthoryear{{Daigne}, {Bo{\v s}njak} \& {Dubus}}{{Daigne}
  et~al.}{2011}]{daigne_11}
{Daigne} F.,  {Bo{\v s}njak} {\v Z}.,    {Dubus} G.,  2011, \aap, 526, A110

\bibitem[\protect\citeauthoryear{{Daughton}}{{Daughton}}{1998}]{daughton_98}
{Daughton} W.,  1998, Journal of Geophysical Research, 103, 29429

\bibitem[\protect\citeauthoryear{{Drenkhahn} \& {Spruit}}{{Drenkhahn} \&
  {Spruit}}{2002}]{spruit_02}
{Drenkhahn} G.,  {Spruit} H.~C.,  2002, \aap, 391, 1141

\bibitem[\protect\citeauthoryear{{Fossati}, {Maraschi}, {Celotti}, {Comastri}
  \& {Ghisellini}}{{Fossati} et~al.}{1998}]{fossatietal_98}
{Fossati} G.,  {Maraschi} L.,  {Celotti} A.,  {Comastri} A.,    {Ghisellini}
  G.,  1998, \mnras, 299, 433

\bibitem[\protect\citeauthoryear{{Gallant}, {Hoshino}, {Langdon}, {Arons} \&
  {Max}}{{Gallant} et~al.}{1992}]{gallant_92}
{Gallant} Y.~A.,  {Hoshino} M.,  {Langdon} A.~B.,  {Arons} J.,    {Max} C.~E.,
  1992, \apj, 391, 73

\bibitem[\protect\citeauthoryear{{Ghisellini} \& {Celotti}}{{Ghisellini} \&
  {Celotti}}{1999}]{ghisellini_99}
{Ghisellini} G.,  {Celotti} A.,  1999, \apjl, 511, L93

\bibitem[\protect\citeauthoryear{{Ghisellini}, {Tavecchio}, {Maraschi},
  {Celotti} \& {Sbarrato}}{{Ghisellini} et~al.}{2014}]{ghisellini_14}
{Ghisellini} G.,  {Tavecchio} F.,  {Maraschi} L.,  {Celotti} A.,    {Sbarrato}
  T.,  2014, \nat, 515, 376

\bibitem[\protect\citeauthoryear{{Giannios}}{{Giannios}}{2006}]{giannios_06}
{Giannios} D.,  2006, \aap, 457, 763

\bibitem[\protect\citeauthoryear{{Giannios}}{{Giannios}}{2010}]{giannios_10}
{Giannios} D.,  2010, \mnras, 408, L46

\bibitem[\protect\citeauthoryear{{Giannios}}{{Giannios}}{2013}]{giannios_13}
{Giannios} D.,  2013, \mnras, 431, 355

\bibitem[\protect\citeauthoryear{{Giannios}, {Mimica} \& {Aloy}}{{Giannios}
  et~al.}{2008}]{giannios_08}
{Giannios} D.,  {Mimica} P.,    {Aloy} M.~A.,  2008, \aap, 478, 747

\bibitem[\protect\citeauthoryear{{Giannios}, {Uzdensky} \&
  {Begelman}}{{Giannios} et~al.}{2009}]{giannios_09}
{Giannios} D.,  {Uzdensky} D.~A.,    {Begelman} M.~C.,  2009, \mnras, 395, L29

\bibitem[\protect\citeauthoryear{{Granot}, {K{\"o}nigl} \& {Piran}}{{Granot}
  et~al.}{2006}]{granot_06}
{Granot} J.,  {K{\"o}nigl} A.,    {Piran} T.,  2006, \mnras, 370, 1946

\bibitem[\protect\citeauthoryear{{Guo}, {Li}, {Daughton} \& {Liu}}{{Guo}
  et~al.}{2014}]{guo_14}
{Guo} F.,  {Li} H.,  {Daughton} W.,    {Liu} Y.-H.,  2014, Physical Review
  Letters, 113, 155005

\bibitem[\protect\citeauthoryear{{Haugb{\o}lle}}{{Haugb{\o}lle}}{2011}]{haugbolle_10}
{Haugb{\o}lle} T.,  2011, \apjl, 739, L42

\bibitem[\protect\citeauthoryear{{Heavens} \& {Drury}}{{Heavens} \&
  {Drury}}{1988}]{heavensdrury_88}
{Heavens} A.~F.,  {Drury} L.~O.,  1988, \mnras, 235, 997

\bibitem[\protect\citeauthoryear{{Homan}, {Kovalev}, {Lister}, {Ros},
  {Kellermann}, {Cohen}, {Vermeulen}, {Zensus} \& {Kadler}}{{Homan}
  et~al.}{2006}]{homan_06}
{Homan} D.~C.,  {Kovalev} Y.~Y.,  {Lister} M.~L.,  {Ros} E.,  {Kellermann}
  K.~I.,  {Cohen} M.~H.,  {Vermeulen} R.~C.,  {Zensus} J.~A.,    {Kadler} M.,
  2006, \apjl, 642, L115

\bibitem[\protect\citeauthoryear{{Hovatta}, {Leitch}, {Homan}, {Wiik},
  {Lister}, {Max-Moerbeck}, {Richards} L. \& {Readhead}}{{Hovatta}
  et~al.}{2013}]{hovatta_13}
{Hovatta} T.,  {Leitch} E.~M.,  {Homan} D.~C.,  {Wiik} K.,  {Lister} M.~L.,
  {Max-Moerbeck} W.,  {Richards} L. J.,    {Readhead} A.~C.~S.,  2013, in
  European Physical Journal Web of Conferences Vol.~61 of European Physical
  Journal Web of Conferences, {Intrinsic brightness temperatures of blazar jets
  at 15 GHz}.
p.~6005

\bibitem[\protect\citeauthoryear{{Jaroschek}, {Lesch} \&
  {Treumann}}{{Jaroschek} et~al.}{2004}]{jaroschek_04}
{Jaroschek} C.~H.,  {Lesch} H.,    {Treumann} R.~A.,  2004, \apjl, 605, L9

\bibitem[\protect\citeauthoryear{{Kagan}, {Milosavljevi{\'c}} \&
  {Spitkovsky}}{{Kagan} et~al.}{2013}]{kagan_13}
{Kagan} D.,  {Milosavljevi{\'c}} M.,    {Spitkovsky} A.,  2013, \apj, 774, 41

\bibitem[\protect\citeauthoryear{{Kagan}, {Sironi}, {Cerutti} \&
  {Giannios}}{{Kagan} et~al.}{2015}]{kagan_15}
{Kagan} D.,  {Sironi} L.,  {Cerutti} B.,    {Giannios} D.,  2015, Space Science
  Reviews

\bibitem[\protect\citeauthoryear{{Kennel} \& {Coroniti}}{{Kennel} \&
  {Coroniti}}{1984a}]{kennelCoroniti84}
{Kennel} C.~F.,  {Coroniti} F.~V.,  1984a, \apj, 283, 694

\bibitem[\protect\citeauthoryear{{Kennel} \& {Coroniti}}{{Kennel} \&
  {Coroniti}}{1984b}]{kennel_84}
{Kennel} C.~F.,  {Coroniti} F.~V.,  1984b, \apj, 283, 710

\bibitem[\protect\citeauthoryear{{Kirk} \& {Heavens}}{{Kirk} \&
  {Heavens}}{1989}]{kirkheavens89}
{Kirk} J.~G.,  {Heavens} A.~F.,  1989, \mnras, 239, 995

\bibitem[\protect\citeauthoryear{{Komissarov}, {Vlahakis}, {K{\"o}nigl} \&
  {Barkov}}{{Komissarov} et~al.}{2009}]{komissarov_09}
{Komissarov} S.~S.,  {Vlahakis} N.,  {K{\"o}nigl} A.,    {Barkov} M.~V.,  2009,
  \mnras, 394, 1182

\bibitem[\protect\citeauthoryear{{Kumar} \& {Barniol Duran}}{{Kumar} \&
  {Barniol Duran}}{2009}]{kumar_09}
{Kumar} P.,  {Barniol Duran} R.,  2009, \mnras, 400, L75

\bibitem[\protect\citeauthoryear{{Kumar} \& {Barniol Duran}}{{Kumar} \&
  {Barniol Duran}}{2010}]{kumar_10}
{Kumar} P.,  {Barniol Duran} R.,  2010, \mnras, 409, 226

\bibitem[\protect\citeauthoryear{{L{\"a}hteenm{\"a}ki}, {Valtaoja} \&
  {Wiik}}{{L{\"a}hteenm{\"a}ki} et~al.}{1999}]{wilk_99}
{L{\"a}hteenm{\"a}ki} A.,  {Valtaoja} E.,    {Wiik} K.,  1999, \apj, 511, 112

\bibitem[\protect\citeauthoryear{{Lemoine}}{{Lemoine}}{2013}]{lemoine_13}
{Lemoine} M.,  2013, \mnras, 428, 845

\bibitem[\protect\citeauthoryear{{Lyubarsky}}{{Lyubarsky}}{2009}]{lyubarsky_09}
{Lyubarsky} Y.,  2009, \apj, 698, 1570

\bibitem[\protect\citeauthoryear{{Lyubarsky}}{{Lyubarsky}}{2005}]{lyubarsky_05}
{Lyubarsky} Y.~E.,  2005, \mnras, 358, 113

\bibitem[\protect\citeauthoryear{{Lyutikov}}{{Lyutikov}}{2006}]{lyutikov_06}
{Lyutikov} M.,  2006, \mnras, 367, 1594

\bibitem[\protect\citeauthoryear{{Lyutikov} \& {Blandford}}{{Lyutikov} \&
  {Blandford}}{2003}]{lyutikov_03}
{Lyutikov} M.,  {Blandford} R.,  2003, ArXiv:astro-ph/0312347

\bibitem[\protect\citeauthoryear{{Martins}, {Fonseca}, {Silva} \&
  {Mori}}{{Martins} et~al.}{2009}]{martins_09}
{Martins} S.~F.,  {Fonseca} R.~A.,  {Silva} L.~O.,    {Mori} W.~B.,  2009,
  \apjl, 695, L189

\bibitem[\protect\citeauthoryear{{Mastichiadis} \& {Kirk}}{{Mastichiadis} \&
  {Kirk}}{1997}]{mastkirk_97}
{Mastichiadis} A.,  {Kirk} J.~G.,  1997, \aap, 320, 19

\bibitem[\protect\citeauthoryear{{McKinney} \& {Uzdensky}}{{McKinney} \&
  {Uzdensky}}{2012}]{mckinney_12}
{McKinney} J.~C.,  {Uzdensky} D.~A.,  2012, \mnras, 419, 573

\bibitem[\protect\citeauthoryear{{Melzani}, {Walder}, {Folini}, {Winisdoerffer}
  \& {Favre}}{{Melzani} et~al.}{2014}]{melzani_14}
{Melzani} M.,  {Walder} R.,  {Folini} D.,  {Winisdoerffer} C.,    {Favre}
  J.~M.,  2014, \aap, 570, A112

\bibitem[\protect\citeauthoryear{{Mimica}, {Giannios} \& {Aloy}}{{Mimica}
  et~al.}{2009}]{mimicaetal09}
{Mimica} P.,  {Giannios} D.,    {Aloy} M.~A.,  2009, \aap, 494, 879

\bibitem[\protect\citeauthoryear{{Nalewajko}, {Giannios}, {Begelman},
  {Uzdensky} \& {Sikora}}{{Nalewajko} et~al.}{2011}]{nalewajko_11}
{Nalewajko} K.,  {Giannios} D.,  {Begelman} M.~C.,  {Uzdensky} D.~A.,
  {Sikora} M.,  2011, \mnras, 413, 333

\bibitem[\protect\citeauthoryear{{Narayan}, {Kumar} \&
  {Tchekhovskoy}}{{Narayan} et~al.}{2011}]{narayan_11}
{Narayan} R.,  {Kumar} P.,    {Tchekhovskoy} A.,  2011, \mnras, 416, 2193

\bibitem[\protect\citeauthoryear{{Narayan} \& {Piran}}{{Narayan} \&
  {Piran}}{2012}]{narayan_12}
{Narayan} R.,  {Piran} T.,  2012, \mnras, 420, 604

\bibitem[\protect\citeauthoryear{{Nava}, {Sironi}, {Ghisellini}, {Celotti} \&
  {Ghirlanda}}{{Nava} et~al.}{2013}]{nava_13}
{Nava} L.,  {Sironi} L.,  {Ghisellini} G.,  {Celotti} A.,    {Ghirlanda} G.,
  2013, \mnras, 433, 2107

\bibitem[\protect\citeauthoryear{{Panaitescu} \& {Kumar}}{{Panaitescu} \&
  {Kumar}}{2002}]{panaitescu02}
{Panaitescu} A.,  {Kumar} P.,  2002, \apj, 571, 779

\bibitem[\protect\citeauthoryear{{Parfrey}, {Giannios} \&
  {Beloborodov}}{{Parfrey} et~al.}{2015}]{parfrey_15}
{Parfrey} K.,  {Giannios} D.,    {Beloborodov} A.~M.,  2015, \mnras, 446, L61

\bibitem[\protect\citeauthoryear{{Pe'er}, {M{\'e}sz{\'a}ros} \& {Rees}}{{Pe'er}
  et~al.}{2006}]{peer_06}
{Pe'er} A.,  {M{\'e}sz{\'a}ros} P.,    {Rees} M.~J.,  2006, \apj, 642, 995

\bibitem[\protect\citeauthoryear{{Petropoulou}, {Dimitrakoudis}, {Padovani},
  {Mastichiadis} \& {Resconi}}{{Petropoulou} et~al.}{2015}]{petro_dim_pado_15}
{Petropoulou} M.,  {Dimitrakoudis} S.,  {Padovani} P.,  {Mastichiadis} A.,
  {Resconi} E.,  2015, ArXiv e-prints

\bibitem[\protect\citeauthoryear{{Readhead}}{{Readhead}}{1994}]{readhead_94}
{Readhead} A.~C.~S.,  1994, \apj, 426, 51

\bibitem[\protect\citeauthoryear{{Rees} \& {Meszaros}}{{Rees} \&
  {Meszaros}}{1994}]{rees_94}
{Rees} M.~J.,  {Meszaros} P.,  1994, \apjl, 430, L93

\bibitem[\protect\citeauthoryear{{Romanova} \& {Lovelace}}{{Romanova} \&
  {Lovelace}}{1992}]{romanova_92}
{Romanova} M.~M.,  {Lovelace} R.~V.~E.,  1992, \aap, 262, 26

\bibitem[\protect\citeauthoryear{{Service}}{{Service}}{1986}]{service86}
{Service} A.~T.,  1986, \apj, 307, 60

\bibitem[\protect\citeauthoryear{{Sikora}, {Stawarz}, {Moderski}, {Nalewajko}
  \& {Madejski}}{{Sikora} et~al.}{2009}]{sikora_stawarz_09}
{Sikora} M.,  {Stawarz} {\L}.,  {Moderski} R.,  {Nalewajko} K.,    {Madejski}
  G.~M.,  2009, \apj, 704, 38

\bibitem[\protect\citeauthoryear{{Sironi} \& {Spitkovsky}}{{Sironi} \&
  {Spitkovsky}}{2009}]{sironi_spitkovsky_09}
{Sironi} L.,  {Spitkovsky} A.,  2009, \apj, 698, 1523

\bibitem[\protect\citeauthoryear{{Sironi} \& {Spitkovsky}}{{Sironi} \&
  {Spitkovsky}}{2011}]{sironi_spitkovsky_11a}
{Sironi} L.,  {Spitkovsky} A.,  2011, \apj, 726, 75

\bibitem[\protect\citeauthoryear{{Sironi} \& {Spitkovsky}}{{Sironi} \&
  {Spitkovsky}}{2014}]{ss_14}
{Sironi} L.,  {Spitkovsky} A.,  2014, \apjl, 783, L21

\bibitem[\protect\citeauthoryear{{Sironi}, {Spitkovsky} \& {Arons}}{{Sironi}
  et~al.}{2013}]{sironi_13}
{Sironi} L.,  {Spitkovsky} A.,    {Arons} J.,  2013, \apj, 771, 54

\bibitem[\protect\citeauthoryear{{Spitkovsky}}{{Spitkovsky}}{2008a}]{spitkovsky_08}
{Spitkovsky} A.,  2008a, \apjl, 673, L39

\bibitem[\protect\citeauthoryear{{Spitkovsky}}{{Spitkovsky}}{2008b}]{spitkovsky_08b}
{Spitkovsky} A.,  2008b, \apjl, 682, L5

\bibitem[\protect\citeauthoryear{{Spruit}, {Daigne} \& {Drenkhahn}}{{Spruit}
  et~al.}{2001}]{spruit_01}
{Spruit} H.~C.,  {Daigne} F.,    {Drenkhahn} G.,  2001, \aap, 369, 694

\bibitem[\protect\citeauthoryear{{Tchekhovskoy}, {McKinney} \&
  {Narayan}}{{Tchekhovskoy} et~al.}{2009}]{sasha_09}
{Tchekhovskoy} A.,  {McKinney} J.~C.,    {Narayan} R.,  2009, \apj, 699, 1789

\bibitem[\protect\citeauthoryear{{Ulrich}, {Maraschi} \& {Urry}}{{Ulrich}
  et~al.}{1997}]{ulrichetal_97}
{Ulrich} M.-H.,  {Maraschi} L.,    {Urry} C.~M.,  1997, \araa, 35, 445

\bibitem[\protect\citeauthoryear{{Werner}, {Uzdensky}, {Cerutti}, {Nalewajko}
  \& {Begelman}}{{Werner} et~al.}{2014}]{werner_14}
{Werner} G.~R.,  {Uzdensky} D.~A.,  {Cerutti} B.,  {Nalewajko} K.,
  {Begelman} M.~C.,  2014, ArXiv e-prints

\bibitem[\protect\citeauthoryear{{Yost}, {Harrison}, {Sari} \& {Frail}}{{Yost}
  et~al.}{2003}]{yost03}
{Yost} S.~A.,  {Harrison} F.~A.,  {Sari} R.,    {Frail} D.~A.,  2003, \apj,
  597, 459

\bibitem[\protect\citeauthoryear{{Zenitani} \& {Hoshino}}{{Zenitani} \&
  {Hoshino}}{2001}]{zenitani_01}
{Zenitani} S.,  {Hoshino} M.,  2001, \apjl, 562, L63

\bibitem[\protect\citeauthoryear{{Zenitani} \& {Hoshino}}{{Zenitani} \&
  {Hoshino}}{2008}]{zenitani_08}
{Zenitani} S.,  {Hoshino} M.,  2008, \apj, 677, 530

\bibitem[\protect\citeauthoryear{{Zhang} \& {Yan}}{{Zhang} \&
  {Yan}}{2011}]{zhang_11}
{Zhang} B.,  {Yan} H.,  2011, \apj, 726, 90

\end{thebibliography}

\end{document}